\documentclass[twocolumn,showpacs,preprintnumbers,amsmath,amssymb,aps,prb,superscriptaddress]{revtex4-2}

\pdfoutput=1
\usepackage[pdftex]{graphicx}
\usepackage[english]{babel}
\usepackage{soul}
\usepackage{cleveref}
\usepackage{bbold}
\usepackage{mathtools}
\usepackage{multirow}
\usepackage{colortbl}
\definecolor{kugray5}{RGB}{224,224,224}
\usepackage[latin1]{inputenc}
\usepackage{diagbox}
\usepackage{mciteplus}
\usepackage{mathrsfs}
\usepackage{amsmath}

\usepackage{dcolumn}
\usepackage{bm}

\usepackage{color}
\usepackage{amstext}
\usepackage{braket}
\usepackage{mathdots}
\usepackage{tikz}
\usepackage{physics}
\usepackage{enumitem}
\usepackage{siunitx}

\usetikzlibrary{matrix,decorations.pathreplacing}

\begin{document}


\title{Flux-mediated effective Su-Schrieffer-Heeger model in an impurity decorated diamond chain}

\author{D. Viedma}
\email{david.viedma@uab.cat}
\affiliation{Departament de F\'isica, Universitat Aut\`onoma de Barcelona, E-08193 Bellaterra, Spain}

\author{A. M. Marques}
\affiliation{Department of physics $\&$ i3N, University of Aveiro, 3810-193 Aveiro, Portugal}

\author{R. G. Dias}
\affiliation{Department of physics $\&$ i3N, University of Aveiro, 3810-193 Aveiro, Portugal}

\author{V. Ahufinger}
\affiliation{Departament de F\'isica, Universitat Aut\`onoma de Barcelona, E-08193 Bellaterra, Spain}




\begin{abstract}
In flat-band systems with non-orthogonal compact localized states (CLSs), onsite perturbations couple neighboring CLSs and generate exponentially-decaying impurity states, whose degree of localization depends on lattice parameters. In this work, a diamond chain with constant magnetic flux per plaquette is decorated with several controlled onsite impurities in a patterned arrangement, generating an effective system that emerges from the flat band. The coupling distribution of the effective system is determined by the relative distance between impurities and the value of the flux, which can be chosen to engineer a wide variety of models. We employ a staggered distribution of impurities that effectively produces the well-known Su-Schrieffer-Heeger model, and show that the topological edge states display an enhanced robustness to non-chiral disorder due to an averaging effect over their extension. Finally, we provide a route to implement the system experimentally using optical waveguides that guide orbital angular momentum (OAM) modes. This work opens the way for the design of topologically protected impurity states in other flat-band systems or physical platforms with non-orthogonal bases.
\end{abstract}

\maketitle

\section{Introduction}
\label{sec:intro}

\begin{figure*}[t]
	\begin{centering}
		\includegraphics[width=1\textwidth]{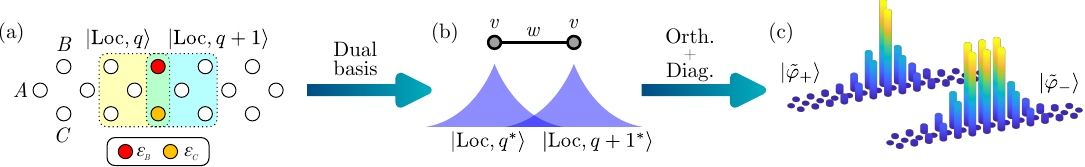} 
		\par\end{centering}
	\caption{Summary of the theoretical description of the effective impurity system. (a) The CLSs of the diamond chain located in the two neighboring plaquettes to the one with onsite impurities $\epsilon_{_B}$ and $\epsilon_{_C}$ (yellow and blue shaded regions) have a direct overlap on these sites. However, the CLSs do not constitute an orthogonal basis, so other states of the FB besides the two CLSs shown in (a) will also have such an overlap when orthogonalizing. To restrict the effective system to two states only, we switch to a dual basis (b) where the impurity operator affects only two localized states that can then be orthogonalized independently to all other states. Finally, diagonalization of the impurity sub-system leads, when reverted back to the site basis of the chain, to the two exponentially-decaying states in (c), whose amplitude distribution depends on the impurity values and the flux per plaquette.}
	\label{fig:1}
\end{figure*}

CLSs \cite{Maimaiti2017} are eigenstates of tight-binding models which possess strictly zero amplitude beyond a finite number of unit cells. These states cannot freely expand throughout the lattice due to destructive interference over different hopping paths, and thus are associated with the absence of transport that occurs in flat-bands (FBs) \cite{Leykam2018}. FB systems possess at least one completely dispersionless band that can arise due to the protection provided by a particular symmetry of the system or the fine-tuning of the parameters of the lattice \cite{Danieli2024}. The CLSs forming the FB are degenerate and generally have nonzero overlap with the adjacent ones \cite{Lopes2014}. Since FBs lack any transport of their own, any perturbation affecting them will have strong effects on the transport properties of the lattice \cite{Danieli2024}.
For example, projecting weak modulated interactions onto gapped FBs \cite{Tovmasyan2013} can lead to the emergence of effective topological two-body subspaces \cite{Pelegri2020,Kuno2020b}.
Both CLSs, and FB systems in general, have garnered significant attention over the last few years, having been implemented in several different physical platforms \cite{Abilio1999,Jacqmin2014,Vicencio2015,Mukherjee2015,Taie2015,Xia2016,Zong2016,Drost2017,Slot2017}. 

We aim to build an effective system with nontrivial topology within a larger system that hosts CLSs, which we take to be the diamond chain with a finite magnetic flux per plaquette, a system that has a gapped zero-energy FB in the energy spectrum. 
By introducing small onsite impurities in a plaquette of the chain, we shift the energies of the CLSs with finite amplitude in the perturbed sites. For finite fluxes, adjacent CLSs become coupled and generate a set of exponentially decaying states whose localization length depends on the flux, and whose shape and phase structure depend on the distribution of impurities \cite{Marques2024}. By decorating the larger lattice with these onsite perturbations, we can build a wide range of effective systems, whose effective coupling parameters can be controlled by tuning the value of the flux and the distance between impurities. To benchmark the method, in this work we imprint a Su-Schrieffer-Heeger (SSH) \cite{Su1979} model onto a diamond chain lattice with constant flux per plaquette. We first briefly describe the theoretical method behind the formation of the effective system, and then focus on the characteristics of the effective system itself, including the appearance of edge states localized around the end of the chain of impurities. 
Additionally, we study the robustness of the effective system against different types of disorder, and show how it displays enhanced protection even against disorder that breaks its chiral symmetry due to the extension of the impurity states. A deeper analysis of the theory behind the generation of the impurity states can be found in Ref.~\cite{Marques2024}.

Finally, we provide a possible route to implement the described system. We require a way to imprint a finite flux per plaquette onto a diamond chain, as well as a way to decorate this chain with controlled onsite impurities at certain sites. In photonics alone, several platforms have been shown to provide a controllable way to introduce these fluxes, such as ring resonators with different optical paths \cite{Hafezi2011,Hafezi2013,Mittal2014,Mittal2016,Viedma2024,Flower2024}, twisted or modulated lattices \cite{Fang2012NP,Longhi2013,Rechtsman2013,Rechtsman2013NP,Plotnik2016,Lumer2019,Parto2019,Russell2023}, fiber loops \cite{Regensburger2011,Regensburger2012,Wimmer2017}, and optical waveguides with multipolar components \cite{Caceres-Aravena2022,Schulz2022,Jiang2023OL} or guiding orbital angular momentum (OAM) modes \cite{Jorg2020,Jiang2023}, among others. We focus on the latter, which additionally allows for a simple method to add the impurities experimentally by slightly modifying parameters such as the writing speed in laser-writing setups \cite{Szameit2010}, or the waveguide width in beam lithography setups \cite{Chen2015}. The coupling between OAM modes carries a phase that can be used to imprint an artificial gauge field on the system \cite{Jorg2020}. To build the diamond, we propose to couple modes with different OAM order $l$ in a zig-zag pattern. Namely, modes with $l=0$ will correspond to the spinal nodes, while modes with $l = \pm 1$ will be split in a synthetic dimension and play the role of top and bottom nodes \cite{Nicolau2023}. In this setup, the relative angle between waveguide pairs will determine the value of the flux, which can be tuned to any desired value.


\begin{figure*}[t]
	\begin{centering}
		\includegraphics[width=1\textwidth]{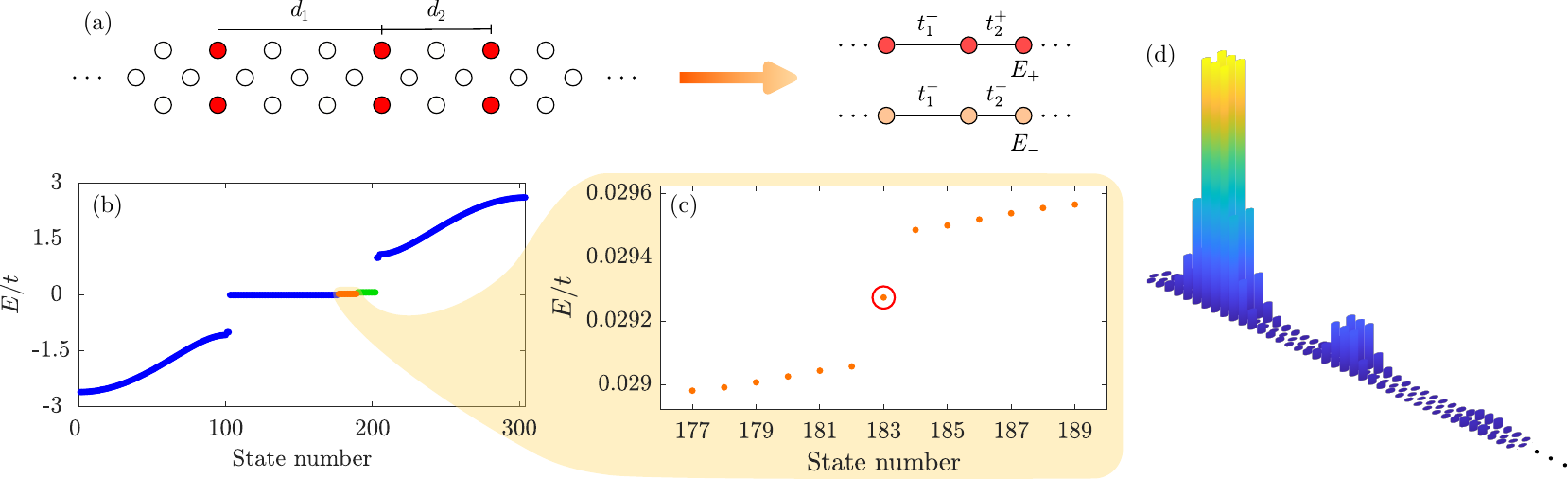} 
		\par\end{centering}
	\caption{(a) A diamond chain decorated with a staggered impurity distribution in the top and bottom sites hosts two effective low-energy SSH sub-systems. (b) Energy spectrum for a diamond chain of $N=101$ unit cells and thirteen impurities $\epsilon_{B} = \epsilon_{C} = \epsilon = 0.1 \, t$, with $\phi = \pi/2$ and relative distances $d_1 = 8$ and $d_2 = 6$ that yield a coupling ratio of $t_1^-/t_2^- = 0.17$. From the FB, two effective SSH systems centered around different energies (orange and green dots) emerge as a consequence of the coupling of impurity states. (c) Spectrum of the lower-energy SSH system (orange dots) highlighted in (b). The state marked with a red circle is its topological edge state. We show in (d) the absolute value of its amplitudes on each site for a segment of the chain.}
	\label{fig:2}
\end{figure*}

\section{Effective impurity systems}

We briefly summarize the theoretical development outlined in Ref.~\cite{Marques2024} and in Fig.~\ref{fig:1}. In particular, we focus on the case of a diamond chain with uniform hoppings $t$ and a non-zero  reduced magnetic flux across each plaquette, $\phi=2\pi\frac{\Phi}{\Phi_0}$, where $\Phi$ is the magnetic flux and $\Phi_0$ the magnetic flux quantum. In this scenario, there exists a basis of non-orthogonal CLSs in the zero-energy FB of the system, each one spanning two plaquettes, see Fig.~\ref{fig:1}(a). Considering the A, B and C sites consecutively at plaquettes $j$ and $j+1$, the $j$-th normalized CLS has the form
\begin{equation}
    \ket{\text{Loc},j}=\frac{1}{2}\big(\cdots,0,1,-e^{-i\frac{\phi}{2}},0,e^{-i\frac{\phi}{2}},-1,\cdots\big)^T,
	\label{eq:locj}
\end{equation}
and is zero elsewhere. As is obvious from expression (\ref{eq:locj}), this CLS will only overlap with the ones at plaquettes $j\pm1$, with strength $S_{j j\pm1} = \braket{\text{Loc},j}{\text{Loc},j+1} = \frac{1}{2} \cos{\frac{\Phi}{2}}=S_{j\pm1 j}$, and thus the overlap matrix of all localized states will be tridiagonal, with main diagonal elements $S_{jj}=1$ from normalization. We now add impurities to the top and bottom sites of a single plaquette at unit cell $q+1$, with $q = N/2$ and $N$ assumed even, of an open diamond chain with $N+1$ unit cells that contains $N$ CLSs in the zero-energy FB. The corresponding impurity operator has the general form
\begin{equation}	\hat{V}=\epsilon_{_B}\ket{B_{q+1}}\bra{B_{q+1}}+\epsilon_{_C}\ket{C_{q+1}}\bra{C_{q+1}},
	\label{eq:impoperator}
\end{equation}
where $\epsilon_{_B}$ and $\epsilon_{_C}$ are local impurity potentials, left as free parameters. From here, one may think of building an effective dimer model out of the two CLSs directly affected by the impurity potentials, i.e., $\ket{\text{Loc},q}$ and $\ket{\text{Loc},q+1}$. However, the basis of CLSs is not orthogonal, and its orthogonalization leads to states that are localized but span across several plaquettes. As a consequence, several other states will also have nonzero weight on the impurity sites, and the effective model will not be restricted to two states. Alternatively, one might work in the dual basis of localized states found via $\ket{\text{Loc},i^*}=\sum\limits_j S_{ij}^{-1}\ket{\text{Loc},j}$ \cite{Marques2024}, where $S_{ij}$ are the elements of the overlap matrix $S$, with $S_{ij}^{-1} \coloneqq [S^{-1}]_{ij}$. Projecting the impurity operator onto this basis leads to:
\begin{eqnarray}
    \hat{V}_{_{FB}} &=& \sum\limits_{n,m=q}^{q+1}\ket{\text{Loc},n^*}V_{nm}\bra{\text{Loc},m^*} \nonumber \\
    &=& v\sum\limits_{n=q}^{q+1}\ket{n^*}\bra{n^*}+ 
	\Big[w\ket{q+1^*}\bra{q^*}+\text{H.c.}\Big], \label{eq:projfb}
\end{eqnarray}
where $V_{nm}=\bra{\text{Loc},n}\hat{V}\ket{\text{Loc},m}$ and we skip the ``Loc'' indicator in the second line for brevity. In the dual basis, $v=\frac{\epsilon_{_B}+\epsilon_{_C}}{4}$ behaves as a local potential on the dual CLSs at $q$ and $q+1$, and $w=\big(\epsilon_{_B}e^{-i\frac{\phi}{2}}+\epsilon_{_C}e^{i\frac{\phi}{2}}\big)/4$ is a hopping term that acts exclusively between these two states, see Fig.~\ref{fig:1}(b). Therefore, one can now independently orthogonalize the two relevant CLSs and work on a reduced subspace of impurity states. Following Ref.~\cite{Marques2024}, one can show that for equal impurities $\epsilon_B=\epsilon_C = \epsilon$, the effective impurity matrix in the new orthogonal basis reads:
\begin{equation}
	\mathbf{\tilde{V}}_{_{FB}}=\frac{\epsilon}{2}\big(\sigma_0 + e^{-\theta}\sigma_x\big), \label{eq:Vmatrix}
\end{equation}
where $\sigma_0$ is the identity and $\sigma_x$ the $x$ Pauli matrix, and where $\theta = \cosh^{-1}{\left(\sec{\frac{\phi}{2}}\right)}$ contains the flux dependence. Diagonalization of this system leads to the two eigenvectors depicted in Fig.~\ref{fig:1}(c), with energies
\begin{equation}
	E_\pm=\frac{\epsilon}{2}\big(1\pm e^{-\theta}\big).
	\label{eq:enereq}
\end{equation}
As illustrated in the right panel of Fig.~\ref{fig:1}, these are exponentially-decaying states which are pulled from the FB, and that will behave as the main sites of an effective system when decorating the lattice with more impurities. The two states lay at different energies within the gap between the FB and the top band, implying that two decoupled copies of the effective system are in fact generated, as sketched in Fig.~\ref{fig:2}(a).


\begin{figure}[t]
	\begin{centering}
		\includegraphics[width=0.90\columnwidth]{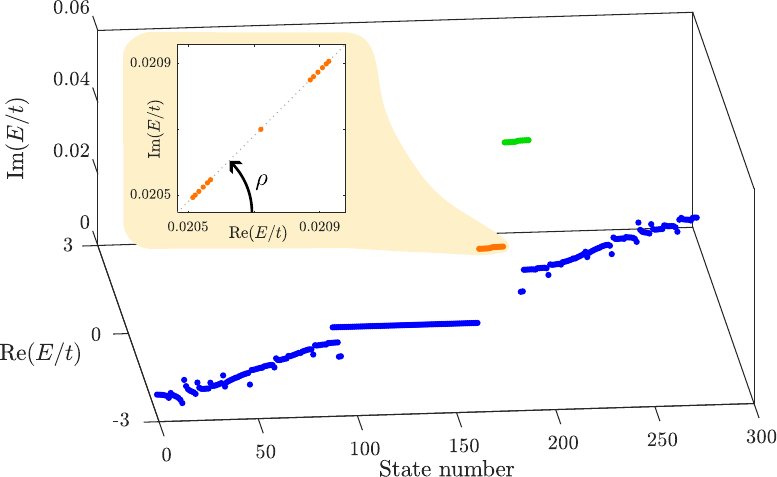} 
		\par\end{centering}
	\caption{Complex energy spectrum for the diamond chain with complex onsite impurities $\epsilon_{B} = \epsilon_{C} = \epsilon e^{i\rho}$, with $\rho = \pi/4$, and the same lattice parameters as Fig.~\ref{fig:2}. The spectra of the effective SSH systems rotate along the complex plane by an angle $\rho$, as can be observed in the inset.
    }
	\label{fig:3}
\end{figure}

\section{Results}
\label{sec:results}

\subsection{Effective SSH impurity chains} \label{sec:results-energy}

We consider a diamond chain of $N = 101$ unit cells with nearest-neighbor coupling $t$ and nonzero flux per plaquette $\phi$, decorated with a series of thirteen impurity pairs in the top and bottom sites, following a staggered pattern as sketched in Fig.~\ref{fig:2}(a). In particular, we use $d_1=8$ and $d_2=6$ plaquettes of separation between impurities, with $d_1$ ($d_2$) at the left (right) end of the chain. The induced impurity states couple between themselves forming an effective system. As described in Supp. Sec.~\ref{sup-sec:imp_coupling}, the coupling between the impurity states depends exponentially on their separation. Therefore, the effective system displays the staggered coupling distribution that is characteristic to the SSH model \cite{Su1979}. We consider the case where the impurities take values $\epsilon_{B} = \epsilon_{C} = 0.1 \, t$ with a flux of $\phi = \pi/2$. 
To avoid cutting the exponentially decaying tails of the edgemost impurity states, which would perturb the ends of the effective system, 13 plaquettes before the first impurity pair are left impurity-free at both ends of the chain.
As we showcase in Fig.~\ref{fig:2}(b), two SSH-like sub-spectra emerge above the FB (orange and green dots). Other choices for the impurity values similarly produce one or two impurity spectra around different energies \cite{Marques2024}. In this manuscript, we focus on the spectrum of lower energy (orange dots) in Fig.~\ref{fig:2}(c), which can be readily recognized to belong to an SSH chain. The highlighted state in the middle of the gap corresponds to the edge state of the effective system, whose amplitudes are shown in Fig.~\ref{fig:2}(d). Interestingly, this edge state is not localized around the edge of the diamond chain, but rather around the edge of the impurity sub-system. 

\begin{figure*}[t]
	\begin{centering}
		\includegraphics[width=1\textwidth]{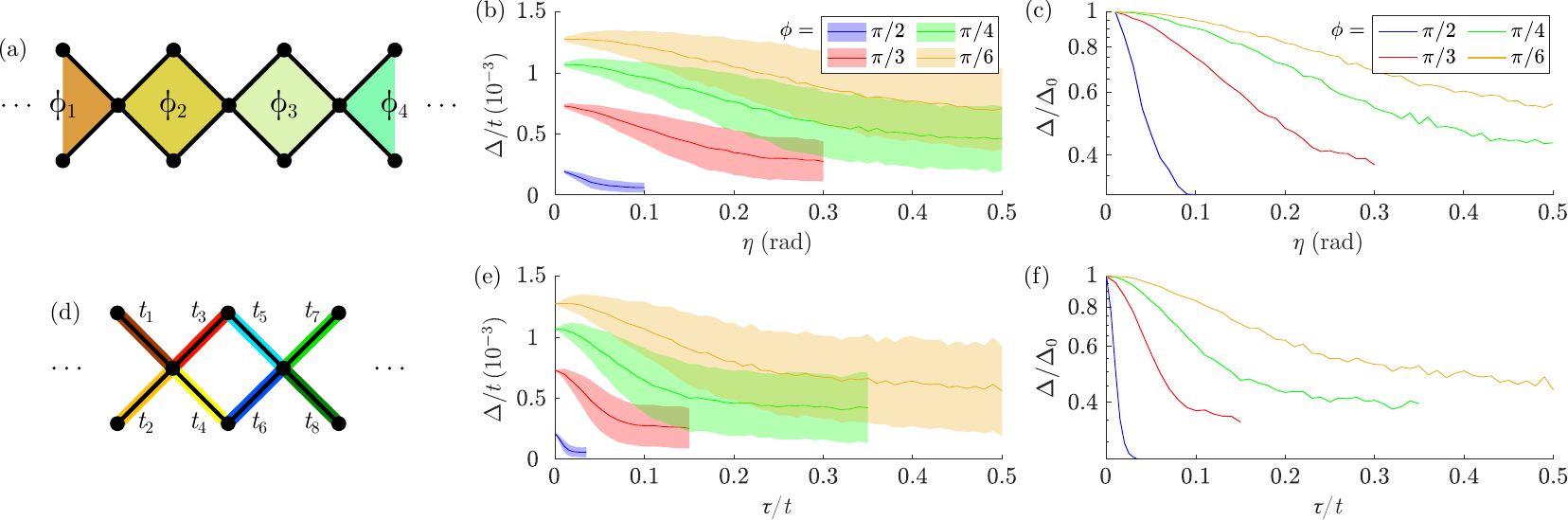} 
		\par\end{centering}
	\caption{Effects of disorder on the spectrum of the effective system of Fig.~\ref{fig:2}(a). (a) Sketch of flux disorder, where each plaquette is threaded by a random flux centered around $\phi_0$. (b) Average gap size $\Delta$ of the edge state of the effective SSH model (solid lines) and standard deviation (shaded region) for increasing flux disorder strengths and different values of $\phi_0$. (c) Disordered gap size $\Delta$ relative to the clean system's gap $\Delta_0$ for different values of $\phi_0$. (d) Sketch of uncorrelated coupling disorder, where each link displays a random coupling sampled uniformly around $t$. (e) Average gap size (solid lines) and standard deviation (shaded regions) for uncorrelated coupling disorder, for the same central flux values $\phi_0$ as in (b). (f) Gap size relative to the clean system's gap for uncorrelated coupling disorder.}
	\label{fig:4}
\end{figure*}

The features of the effective systems may be controlled by tuning the parameter values of the diamond chain. By choosing an appropriate distribution for the impurities, one may design a wide range of one-dimensional (1D) models. 
Nonetheless, the flux through each plaquette remains a key property, as it affects both the energies and the couplings of the effective system, as can be seen from (\ref{eq:Vmatrix}). In particular, the extension of the impurity states grows for weaker flux, as does the coupling between states localized around adjacent impurity pairs and the gap of the effective system. This fact is proved in Supp. Sec.~\ref{sup-sec:imp_coupling}, where the coupling strength is plotted for varying fluxes and different impurity separations. The changes in the extension of the states with the flux are further showcased in Ref.~\cite{Marques2024}.

Most notably, the coupling between impurity states is not limited only to real values and can be pushed into the complex plane by including complex onsite potentials. Namely, one can impose $\epsilon = |\epsilon| e^{i\rho}$, which has two main consequences in the spectrum, as we showcase in Fig.~\ref{fig:3} for $\rho = \pi/4$. Firstly, all eigenstates in the conductive bands with finite weight on the impurity sites pick up a small imaginary component in their energies. This effect, however, is nullified when the impurities are balanced, i.e. $\epsilon_B = - \epsilon_C$. Secondly, and more interestingly, the spectra of the effective SSH systems rotate along the complex plane according to the angle $\rho$, as can be seen in the inset in Fig.~\ref{fig:3}. By virtue of this rotation, one can readily observe that the effective Hamiltonian of each SSH becomes $\tilde{H}_{\text{eff}} = e^{i\rho}H_{\text{eff}}$ compared to the case of Fig.~\ref{fig:2}, implying that they are now \textit{non-Hermitian}. This effect could open up the possibility of experimentally exploring non-Hermitian couplings in FB systems by employing lossy impurities, where the coupling strength is determined by the flux per plaquette and the angle of rotation $\rho$ by the included losses. 

\subsection{Disorder} \label{sec:results-disorder}

Due to the dependence of the extension of the impurity states on the flux, a disorder on the flux translates into disorder in both the onsite energies and the couplings of the effective system, and thus breaks the chiral symmetry of the effective SSH model. We introduce a random flux disorder of strength $\eta$ in each plaquette, $\phi_j = \phi_0 + \eta_j$, sampled from a uniform distribution $\eta_j \in \left[-\eta/2,\eta/2\right]$ as sketched in Fig.~\ref{fig:4}(a). We explore the consequences of this disorder by checking the variation of the edge state gap $\Delta = \text{min}(|E_q - E_{q\pm1}|)$ for the edge state $q$ of the SSH system in Fig.~\ref{fig:2}(c) as the perturbation is increased. We plot in Fig.~\ref{fig:4}(b) the size of this gap and its standard deviation for different values of the central flux $\phi_0$, averaged over 1000 random realizations for each $\eta$, in steps of $\Delta\eta=\SI{0.01}{\radian}$. We see that disorder distorts the effective system, reducing the gap on average for increasing flux disorder. In general, however, for a weak $\phi_0$ the system displays a larger gap that is also more resilient to disorder compared to fluxes closer to $\pi$. This is made evident in Fig.~\ref{fig:4}(c), where the average gap is compared to the one for the pristine system, $\Delta_0$, for different values of $\phi$. The plot shows how weaker fluxes display a slower decay, even for values of disorder of the same order as $\phi_0$ itself. This behavior is closely related to the extension of the impurity states. 
Although flux disorder breaks chiral symmetry, the impurity states feel the effects of an averaged disorder over their whole characteristic length \cite{Munoz2018}, which tends to the effective restoration of chiral symmetry. 
This average effect becomes stronger for longer extensions, implying an overall weaker effect of the disorder, hence explaining the higher robustness against it as one reduces the flux \cite{Marques2024}. The flux-dependent extension of the impurity states thus becomes a pivotal property of the system.

On the other hand, when disorder in the couplings on the diamond chain is introduced in a correlated manner, as detailed in Supplementary Section~\ref{sup-sec:disorder}, this translates into off-diagonal disorder in the effective system, against which it is protected. In contrast, completely uncorrelated coupling disorder, as sketched in Fig.~\ref{fig:4}(d), affects the amplitude distribution of the CLSs themselves as well as their overlap \cite{Marques2024}, thus affecting the effective impurity system in a similar manner to flux disorder. We prove this fact by including a random coupling disorder in each link such that $t_j = t + \tau_j$, for $\tau_j \in \left[-\tau/2,\tau/2\right]$. In Figs.~\ref{fig:4}(e) and (f), we plot the average gap size and its standard deviation for different values of $\phi_0$ and its comparison with the one on the pristine system, respectively. Following the trend depicted in Figs.~\ref{fig:4}(b) and (c) for flux disorder, we observe how uncorrelated coupling disorder heavily affects the spectrum and the gap of the effective system, but is partially resisted on average for low fluxes even for disorder strengths up to 50\% of the value of $t$. 
In contrast, and as we showcase in Supplementary Section~\ref{sup-sec:disorder}, the effective system is protected against correlated disorder due to its chiral symmetry. 
Results for the other non-highlighted subsystem are similar to those depicted here. Nonetheless, when comparing the two systems, the lower-energy one displays a higher robustness to disorder owing to the higher extension of its states that enhances the disorder averaging effect \cite{Marques2024}.

\begin{figure*}[t]
	\begin{centering}
		\includegraphics[width=0.90\textwidth]{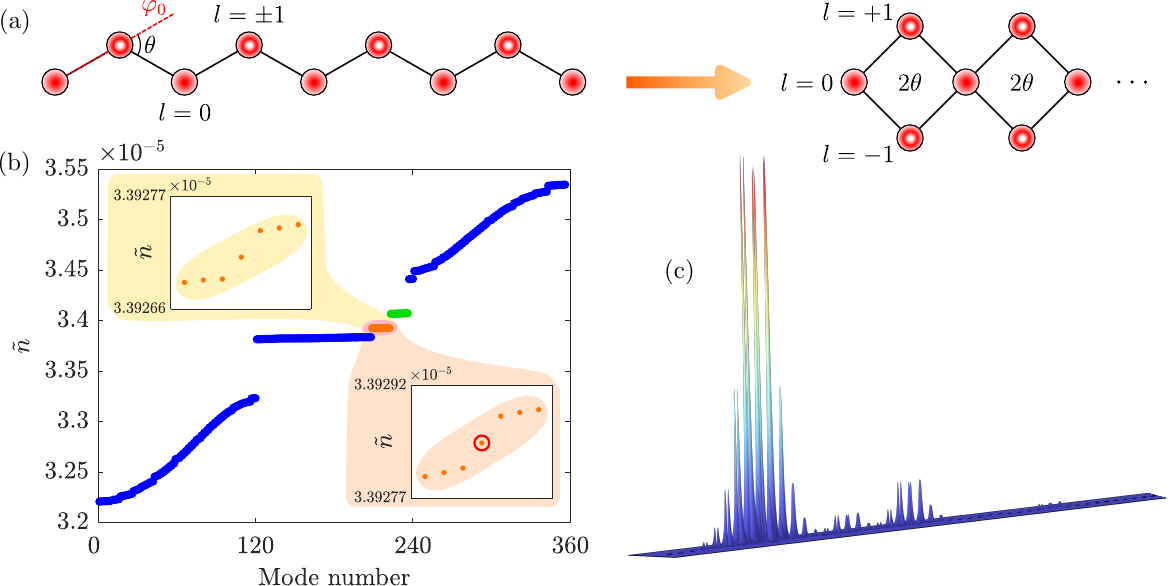} 
		\par\end{centering}
	\caption{Implementation of the effective SSH model in a lattice of waveguides hosting OAM modes. (a) Coupled optical waveguides forming a zig-zag chain with alternating $l=0$ and $l=\pm1$ OAM charges and a staggered relative angle $\theta$ with respect to the reference axis $\varphi_0$, which behaves as diamond chain lattice with a flux per plaquette of $2\theta$. The $l=\pm1$ modes can be split in a synthetic dimension that forms the top and bottom sites of the diamond. (b) Spectrum of effective mode indices with respect to the cladding index, $\tilde{n} = n_{ef} - n_0$, for a waveguide lattice of $N=59$ unit cells, a relative angle of $\theta = \pi/4$ ($\phi = \pi/2$) and 13 impurity pairs in a staggered pattern $d_1 = 8$ and $d_2 = 6$. The insets highlight the lower-energy SSH sub-system (orange dots). Since each waveguide can host two orthogonal mode polarizations, the eigensolver picks up two orthogonal spectra for each SSH system and the spectrum appears doubled. (c) Electric field norms on each waveguide for the edge state circled in red in (b).}
	\label{fig:5}
\end{figure*}

\section{Implementation} \label{sec:implementation}

We consider a zig-zag chain of optical waveguides with a relative distance of $d$ and a staggered relative angle of $\theta$, as depicted in Fig.~\ref{fig:5}(a). These will guide orbital angular momentum (OAM) modes of the form
\begin{equation}
    \Psi^{\pm l} (r,\varphi,z) = \psi^{\pm l}(r) e^{\pm i l\left(\varphi-\varphi_0\right)} e^{-i\beta_l z}, \label{eq:OAM}
\end{equation}
where $l$ is the OAM charge, $(r,\varphi,z)$ are the cylindrical coordinates centered at each waveguide, $\psi^{\pm l}(r)$ the radial mode profile, $\varphi_0$ an arbitrary phase origin and $\beta_l$ the propagation constant of the mode. According to the modes that are guided in each case, we will consider two types of waveguides. The first set [lower row in Fig.~\ref{fig:5}(a)] will guide only the fundamental $l=0$ mode, while the second set [upper row in Fig.~\ref{fig:5}(a)] will also guide the $l=\pm1$ modes. By tuning the refractive index contrast of both sets, one can achieve phase matching between the aforementioned $l=0$ and $l=\pm1$ modes. In such a scenario, this subset of modes forms an effective diamond chain model by considering the mode circulations as a synthetic dimension. The $l=0$ modes become the central sites and the $l=\pm1$ the top and bottom sites of the diamond, respectively, as shown in the right-hand side of Fig.~\ref{fig:5}(a). Furthermore, it is known that the coupling between OAM modes picks up a phase according to their momentum charge and their circulation, as well as the coupling angle $\theta$ with respect to an arbitrary direction $\varphi_0$ [see Fig.~\ref{fig:5}(a)], i.e., $t_{l_1,l_2} = t e^{i(l_1-l_2)\theta}$ \cite{Polo2016,Pelegri2019,Pelegri2019b}. Considering our geometry, and that we are coupling $l_1 = \pm1$ and $l_2 = 0$ modes, this implies a flux per plaquette of twice the geometric angle $2\theta$ throughout the diamond chain. Therefore, any arbitrary value for the flux can be generated by tuning this angle. The implementation of the system in Fig.~\ref{fig:2}(a) is completed by perturbing the refractive index of the $l=\pm1$ waveguides. Since the two OAM components that form the top and bottom sites in the diamond come from the same physical waveguide, we are limited to a symmetric impurity $\epsilon_B = \epsilon_C = \epsilon$ distribution in the effective system.

For our numerical simulations, we consider cylindrical waveguides of radius $R=\SI{12}{\micro\meter}$ separated by a relative distance $d = \SI{55}{\micro\meter}$ and with a cladding refractive index of $n = \num{1.48}$. The $l=0$ waveguides display a contrast of $\Delta n_0 = \num{1e-4}$ and the matching contrast for $l=\pm1$ waveguides is computed to be $\Delta n_1 = \num{2.579e-4}$ for a wavelength of $\lambda = \SI{730}{\nano\meter}$. We impose small impurities on top of some $l=\pm 1$ waveguides by adding an additional contrast of $\Delta n_{imp} = \num{6e-7}$ at alternating distances of 8 and 6 unit cells to achieve the staggering, and choose a relative angle of $\theta = \pi/4$ equivalent to a flux value of $\phi = \pi/2$. Using these parameter values, we simulate a chain of $N=59$ unit cells using the commercial finite-element solver COMSOL Multiphysics. As depicted in Fig.~\ref{fig:5}(b), we obtain the eigenmodes of the structure, closely matching those obtained in the tight-binding model described in Section~\ref{sec:results}. The insets of the figure show that we obtain a doubled spectrum, which is due to the orthogonal mode polarizations that can be guided by the waveguide structure. These have similar mode profiles but different projections $(E_x,E_y,E_z)$, and are both picked up by the eigensolver. We showcase in Fig.~\ref{fig:5}(c) the amplitudes $|E|$ in the waveguide system of the highlighted edge state in the bottom right inset of Fig.~\ref{fig:5}(b).

The simulations display a very good agreement with the theoretical results. However, there exist two possible sources of discrepancy with the original model in the form of additional couplings. For the limiting case of $\theta = \pi/2$, and thus a flux of $\pi$, the next-nearest neighbor (NNN) distance between waveguides becomes as short as $\sqrt{2}d$. Even then, considering their exponentially decaying nature, the NNN couplings are usually neglected \cite{Jorg2020}. We show in Supplementary Section~\ref{sup-sec:OAM_coupling} that they are an order of magnitude lower than the main coupling $t$, but nonetheless they perturb the bands of the system and cause some states of the flat band to tilt to lower energies. 
Since our method relies on pulling CLSs from this flat band, we limit ourselves to angles smaller than $\pi/4$ (and thus fluxes smaller than $\pi/2$) where these couplings are already two orders of magnitude lower than $t$, and therefore can be safely ignored. Another effect of interest is the appearance of self-couplings within the $l=\pm1$ waveguides due to the breaking of the cylindrical symmetry by the presence of other nearby waveguides \cite{Polo2016}. In Supplementary Section~\ref{sup-sec:OAM_coupling} we prove how these self-couplings are around two orders of magnitude lower than $t$ for our parameter values, so they also have a negligible effect on our system.

\section{Conclusions}

We have demonstrated a method to engineer an arbitrary system from the CLSs of a FB lattice, and whose parameters are controlled via the flux that threads the plaquettes of the original system. By decorating the FB lattice with onsite impurities, CLSs can be made to couple, giving rise to an effective system in the subspace of exponentially decaying impurity states. An appropriate choice for the impurity positions leads to a manifestation of non-trivial topology and the appearance of edge states.
To exemplify the method, we imprint an SSH model on top of a diamond chain lattice with nonzero flux per plaquette. By alternating the relative distance between impurities, the characteristic staggered coupling distribution of the SSH model is achieved. We then study the behavior of the effective system when disorder of two kinds is introduced in the system. When correlated coupling disorder is applied, and thus chiral symmetry is preserved, the effective system is immune to it. On the other hand, when flux disorder or uncorrelated coupling disorder is introduced, chiral symmetry is no longer satisfied. Nonetheless, owing to the large spatial extension of the impurity states, the disorder is averaged out over their characteristic lengths, which in turn implies a lower distortion of the energy spectrum of the effective system \cite{Munoz2018}. This effect is also amplified at lower fluxes, where the extension of the states is larger \cite{Marques2024}.

Additionally, we provide a route for an experimental implementation of the proposed system using optical waveguides guiding light with OAM. The coupling of different OAM modes introduces a phase component in the couplings by controlling the geometric angle between waveguides. This phase serves as an artificial gauge field that provides the necessary flux to the system. Moreover, the different circulations of $l = \pm 1$ OAM modes are used as a synthetic dimension and translated as the top and bottom sites of the diamond chain, while $l=0$ modes form the central sites. The onsite impurities can be included by manipulating the propagation constant of the corresponding waveguides. The proposed platform gives complete freedom over the choice of the flux per plaquette, closely replicating the diamond chain for fluxes lower than $\pi/2$ and with minor band distortions for fluxes nearing $\pi$.

The method for generating effective models by impurity decorating flat band systems with non-orthogonal bases can be readily generalized to higher-dimensional systems. 
Furthermore, and as exemplified here, non-Hermitian couplings are also easily achievable by including onsite gains and losses, opening a new FB-based venue for the study of non-Hermitian physics.
Aside from optical waveguides, other platforms capable of controlling the phase of the couplings, such as ring resonators, could be well suited for an alternative experimental implementation.

Note added: During the last stages of writing the manuscript, an independent proposal to implement an impurity-free diamond chain in optical waveguides using OAM was published in Ref.~\cite{Wang2024}.

\begin{acknowledgments}

D.V. and V.A. acknowledge financial support from the Spanish State Research Agency AEI (contract No. PID2020-118153GBI00/AEI/10.13039/501100011033) and Generalitat de Catalunya (Contract No. SGR2021-00138).
A.M.M. and R.G.D. developed their work within the scope of Portuguese Institute for Nanostructures, Nanomodelling and Nanofabrication (i3N) Projects No. UIDB/50025/2020, No. UIDP/50025/2020, and No. LA/P/0037/2020, financed by national funds through the Funda\c{c}\~ao para a Ci\^encia e Tecnologia (FCT) and the Minist\'erio da Educa\c{c}\~ao e Ci\^encia (MEC) of Portugal.
A.M.M. acknowledges financial support from i3N through the work Contract No. CDL-CTTRI-46-SGRH/2022.

\end{acknowledgments}

\section{Supplementary}

\subsection{Coupling between impurity states} \label{sup-sec:imp_coupling}

\begin{figure}[h]
	\begin{centering}
		\includegraphics[width=0.90\columnwidth]{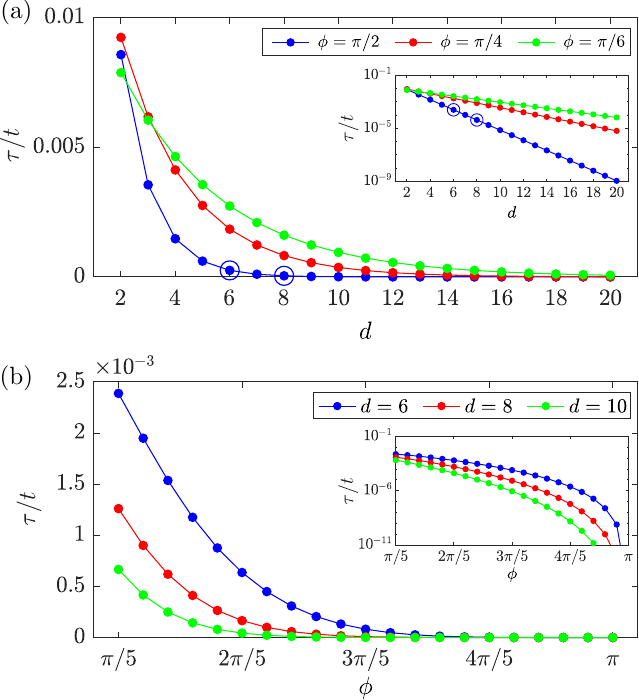} 
		\par\end{centering}
	\caption{(a) Coupling strength between two sets of impurities with $\epsilon_B = \epsilon_C = 0.1\, t$ with respect to the number of plaquettes separating them for different flux values. The inset shows the same curves in a logarithmic scale. The coupling values chosen for Fig.~\ref{fig:2}(b) in the main text, namely those for $d_1=6$ and $d_2=8$ for $\phi=\frac{\pi}{2}$, are highlighted with blue circles. (b) Coupling strength for the same impurity values, this time for varying fluxes and fixed impurity distances. The higher extension of the states for lower fluxes is reflected by the increased coupling strengths.}
	\label{fig:s1}
\end{figure}

To check the exact dependence of the coupling between different impurity states, we place two sets of equal impurities on different plaquettes in the diamond chain. To extract the coupling, we check the energy gap between the two induced impurity states with lower energy as we change the relative distance, and define the coupling as
\begin{equation}
    t = \frac{1}{2}\left(E_{+}-E_{-}\right),
\end{equation}
with $E_{+}$ ($E_{-}$) the energy of the higher (lower) state. We plot the obtained coupling dependence with the distance between impurities in Fig.~\ref{fig:s1}(a), where we observe that it displays an exponential dependence on their separation, confirmed also by the inset. We also see that the decaying of the curves is slower for smaller fluxes, in agreement with the larger extension that the impurity states present \cite{Marques2024}. This is further proved in Fig.~\ref{fig:s1}(b), where we now plot the coupling with respect to the flux for different separations. Since the coupling is larger for lower fluxes, the extension of the states is necessarily larger as well. The curves in Fig.~\ref{fig:s1} allow us to build a map between couplings and distances, which can also be tuned by controlling the flux on the lattice. Based on it, any coupling distribution can be engineered. The plot for the set of higher-energy impurity states presents very similar results as those depicted in Fig.~\ref{fig:s1}. This may seem counter-intuitive, considering the fact that the lower energy states possess a comparatively larger extension. The amplitudes of higher- and lower-energy impurity states for different fluxes can be found in \cite{Marques2024}. When numerically computing the coupling, the different behavior of the central peak at the plaquette hosting the impurities as well as the exponential amplitude tails in both cases seems to lead to approximately equal results.


\subsection{Correlated coupling disorder} \label{sup-sec:disorder}

\begin{figure}[t]
	\begin{centering}
		\includegraphics[width=0.90\columnwidth]{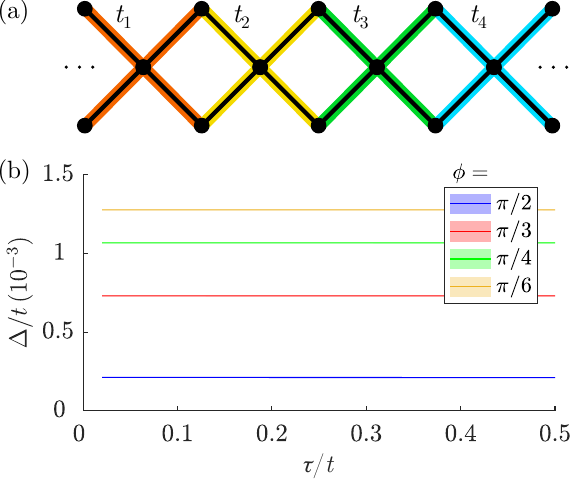} 
		\par\end{centering}
	\caption{Effects of correlated coupling disorder on the spectrum of the effective system. (a) Sketch of the considered disorder, introduced in quartets as indicated by the different link colors. (b) Average gap size $\Delta$ of the edge state of the effective SSH model (solid lines) and standard deviation (shaded region) for different flux values. In this scale, the shaded region is too thin to be observed.}
	\label{fig:s2}
\end{figure}

As described in the main text, coupling disorder only behaves as chiral for the effective system when introduced in a correlated manner. As CLSs occupy two plaquettes for fluxes between 0 and $\pi$, any coupling disorder within the span of each CLS will affect their amplitudes and energies, hence breaking the chiral symmetry of the effective system. For the disorder to be chiral, it has to be introduced in hopping quartets \cite{Viedma2024} as highlighted in Fig.~\ref{fig:s2}(a), so that it may affect the coupling between impurity states without distorting the CLSs themselves. We plot in Fig.~\ref{fig:s2}(b) the effects of this disorder on the gap size of the effective system. Comparing with Fig.~\ref{fig:4} in the main text, we observe that the edge state gap is immune to correlated disorder even for $\phi = \pi/2$.

\begin{figure}[t]
	\begin{centering}
		\includegraphics[width=0.90\columnwidth]{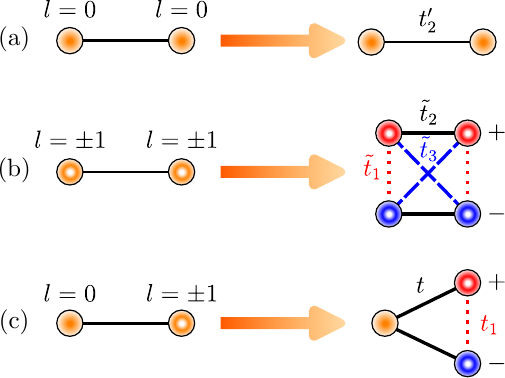} 
		\par\end{centering}
	\caption{Coupling scheme between waveguides with different OAM modal content. The left side corresponds to the physical waveguides, whereas in the right side different OAM circulations are represented separately.}
	\label{fig:s3}
\end{figure}
\begin{figure*}[tp]
	\begin{centering}
		\includegraphics[width=0.90\textwidth]{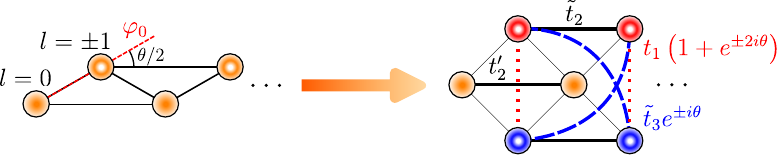} 
		\par\end{centering}
	\caption{Full coupling scheme in the effective diamond chain featured in the main text, including the phases induced by the geometrical angle. The NN couplings are also sketched for clarity. Cross-circulation couplings in the $l = \pm1$ waveguides due to the nearest $l=\pm1$ waveguides, $\tilde{t}_1$, are much smaller than $t_1$ and are thus not included.}
	\label{fig:s4}
\end{figure*}

\subsection{Coupling between OAM modes} \label{sup-sec:OAM_coupling}

As described in Ref.~\cite{Polo2016}, symmetry considerations reduce the number of independent components in the analysis of the coupling between OAM modes. In addition to the nearest-neighbor couplings discussed in the main text, one may also consider the effect of longer-range couplings between the closest waveguides of the same type. Hence, for our purposes, it is enough to consider the following three scenarios: (a) $l=0 \leftrightarrow l=0$, (b) $l=\pm 1 \leftrightarrow l=\pm1$ and (c) $l=0 \leftrightarrow l=\pm 1$, as sketched in Fig.~\ref{fig:s3}. All cases can be described via at most three independent couplings: between different OAM components in the same waveguide, $t_1$, between the same OAM component in different waveguides, $t_2$, and between different components in different waveguides, $t_3$. These depend on the overlap integral between the relevant modes, and $t_1$ and $t_3$ are also subject to picking up a phase depending on their OAM charge and the relative angle with respect to an arbitrary origin $\varphi_0$ \cite{Polo2016}. The first coupling, $t_1$, appears due to the breaking of the cylindrical symmetry of the waveguide mode due to the presence of the second waveguide. We specify all possible next-nearest-neighbor (NNN) components that appear in our system in Fig.~\ref{fig:s4}, with their corresponding phases. Note that the geometrical angle between $l=\pm 1$ waveguides is half as large as the one between $l=0$ and $l=\pm 1$.
\begin{figure*}[tp]
	\begin{centering}
		\includegraphics[width=0.90\textwidth]{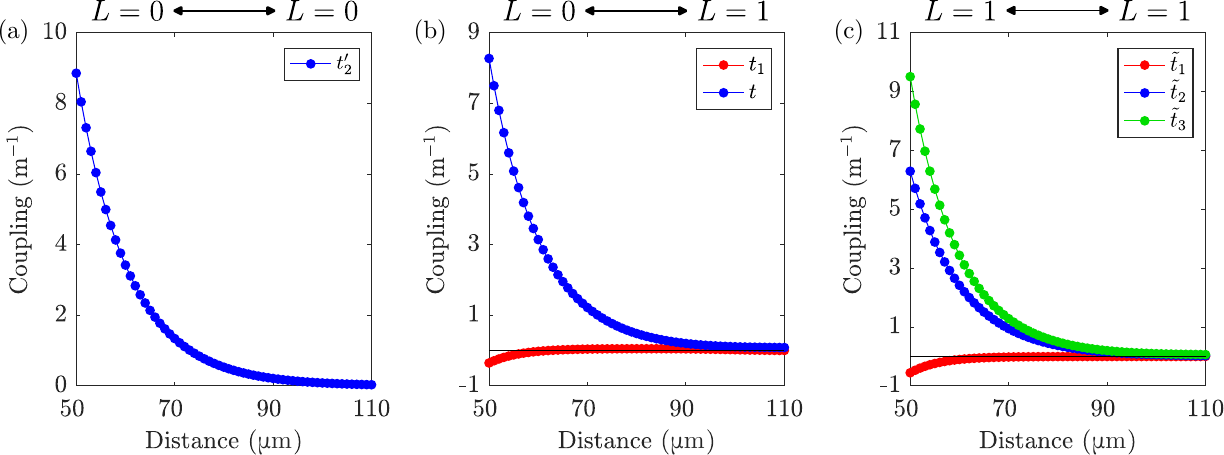} 
		\par\end{centering}
	\caption{Coupling strengths between waveguides with different OAM modal content at increasing relative distances, for the waveguide parameter values included in the main text. The $t_1$ coupling is at least an order of magnitude lower than the rest at all distances, both in (b) and (c).}
	\label{fig:s5}
\end{figure*}
\begin{figure*}[tp]
	\begin{centering}
		\includegraphics[width=0.90\textwidth]{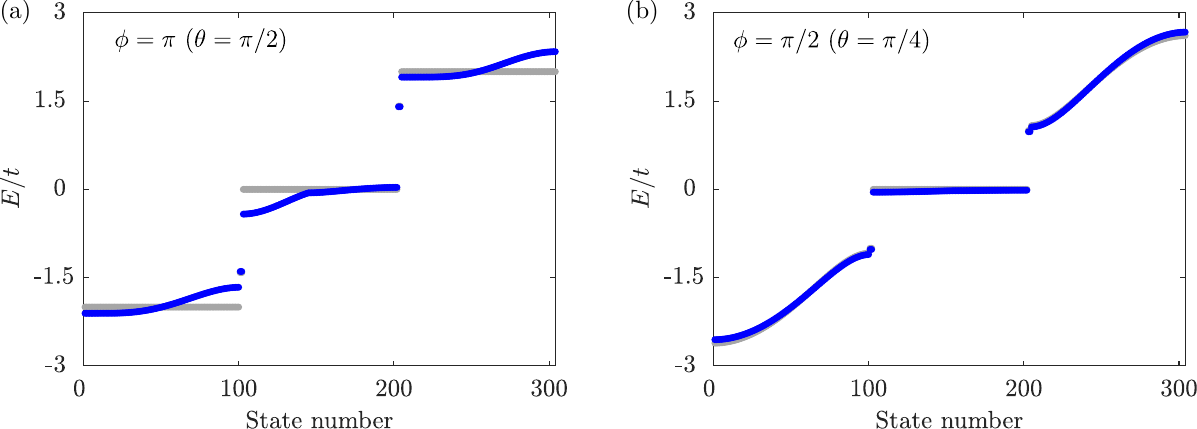} 
		\par\end{centering}
	\caption{Energy spectrum of the diamond chain for (a) $\phi = \pi$ ($\theta = \pi/2$) and (b) $\phi = \pi/2$ ($\theta = \pi/4$), using the coupling strengths obtained from the waveguide simulations in Fig.~\ref{fig:s5} at $d = \SI{55}{\micro\meter}$ and for $N=101$ plaquettes. The spectrum in blue corresponds to a system with both a cross-circulation coupling and the NNN couplings included in Table~\ref{tab:couplings}, whereas the gray spectrum corresponds to a system with only NN couplings.}
	\label{fig:s6}
\end{figure*}
We now describe how to compute these couplings from the eigenvalues of the tight-binding model generated by the couplings in Fig.~\ref{fig:s3}(a)-(c). For case (a), the Hamiltonian is given by
\begin{equation}
    H_{l=0} = \begin{pmatrix}
        \beta_0 & t_2' \\
        t_2' & \beta_0
    \end{pmatrix},\label{H_l0}
\end{equation}
whose eigenvalues are $\lambda_{-} = \beta_0 - t_2'$ and $\lambda_{+} =\beta_0 + t_2'$, and thus
\begin{eqnarray}
    t_2' = \frac{1}{2}\left(\lambda_+ - \lambda_-\right). \label{sup-eq:t0}
\end{eqnarray}
The Hamiltonian for case (b) is written as
\begin{equation}
    H_{l=1} = \begin{pmatrix}
        \beta_1 & \tilde{t}_1 & \tilde{t}_2 & \tilde{t}_3\\
        \tilde{t}_1 & \beta_1 & \tilde{t}_3 & \tilde{t}_2\\
        \tilde{t}_2 & \tilde{t}_3 & \beta_1 & \tilde{t}_1\\
        \tilde{t}_3 & \tilde{t}_2 & \tilde{t}_1 & \beta_1
    \end{pmatrix},\label{H_l1}
\end{equation}
whose eigenvalues are
\begin{align}
    \lambda_1 &= \beta_1 + \tilde{t}_1 - \tilde{t}_2 - \tilde{t}_3, \nonumber \\
    \lambda_2 &= \beta_1 - \tilde{t}_1 + \tilde{t}_2 - \tilde{t}_3, \nonumber \\
    \lambda_3 &= \beta_1 - \tilde{t}_1 - \tilde{t}_2 + \tilde{t}_3, \nonumber \\
    \lambda_4 &= \beta_1 + \tilde{t}_1 + \tilde{t}_2 + \tilde{t}_3,
\end{align}
and thus the couplings can be determined as
\begin{align}
    \tilde{t}_1 &= \lambda_1 - \lambda_2 - \lambda_3 + \lambda_4, \nonumber \\
    \tilde{t}_2 &= -\lambda_1 + \lambda_2 - \lambda_3 + \lambda_4, \nonumber \\
    \tilde{t}_3 &= -\lambda_1 - \lambda_2 + \lambda_3 + \lambda_4. \label{sup-eq:t_11}
\end{align}
Finally, for case (c) we have
\begin{equation}
    H_{l=0,1} = \begin{pmatrix}
        \beta_2 & t & t_1\\
        t & \beta_2 & t_1\\
        t & t_1 & \beta_2
    \end{pmatrix},\label{H_l0,1}
\end{equation}
for which the eigenvalues are
\begin{align}
    \lambda_1 &= \beta_2 - t_1, \nonumber \\
    \lambda_2 &= \beta_2 + \frac{1}{2}\left(t_1 - \sqrt{t_1^2+8t^2}\right), \nonumber \\
    \lambda_3 &= \beta_2 + \frac{1}{2}\left(t_1 + \sqrt{t_1^2+8t^2}\right),
\end{align}
and the couplings read
\begin{align}
    t_1 &= \frac{1}{3}\left(-2\lambda_1 + \lambda_2 + \lambda_3\right), \nonumber \\
    t &= \sqrt{\frac{(\lambda_3-\lambda_2)^2-(-2\lambda_1 + \lambda_2 + \lambda_3)^2/9}{8}}. \label{sup-eq:t1}
\end{align}

We perform finite-element simulations of the waveguides described in the main text at different distances and compute the coupling strengths using expressions (\ref{sup-eq:t0}), (\ref{sup-eq:t_11}) and (\ref{sup-eq:t1}), as displayed in Fig.~\ref{fig:s5}. Using the computed values, we can estimate the importance of the additional couplings not taken into account in the main text, and thus the deviation of the implementation from a pure diamond chain. For a relative distance of $d = \SI{55}{\micro\meter}$, the nearest-neighbor (NN) coupling has a value of $t = \SI{5.08}{m^{-1}}$ and the cross-circulation coupling at the same distance is $t_1 = -\SI{0.12}{m^{-1}} = -0.024 \, t$. Depending on the chosen angle $\theta$, the NNN distance $d'$ will vary. We summarize the NNN couplings for $\theta = \pi/2$ ($d' = \SI{77.8}{\micro\meter}$) and $\theta = \pi/4$ ($d' = \SI{101.64}{\micro\meter}$) in Table~\ref{tab:couplings}. We do not include the cross-circulation coupling $\tilde{t}_1$ as it is at least an order of magnitude lower than the others at all distances.
\begin{table}[h]
    \begin{tabular}{c|c|c|}
        \cline{2-3}
                               & $\phi = \pi$ & $\phi = \pi/2$ \\ \hline
        \multicolumn{1}{|c|}{$t'_2 (l=0\leftrightarrow l=0)$}  & 12.82\% & 2.58\% \\ \hline
        \multicolumn{1}{|c|}{$\tilde{t}_2 (l=\pm1\leftrightarrow l=\pm1)$}  & 9.04\% & 1.82\% \\ \hline
        \multicolumn{1}{|c|}{$\tilde{t}_3 (l=\pm1\leftrightarrow l=\mp1)$}  & 12\% & 2.58\% \\ \hline
    \end{tabular}
    \caption{NNN coupling strength values for different flux values, relative to the value for the NN coupling at $d = \SI{55}{\micro\meter}$.}
    \label{tab:couplings}
\end{table}

With the coupling strengths in Table~\ref{tab:couplings}, we simulate the tight-binding model of a diamond chain model with NNN and cross-circulation couplings, see Fig.~\ref{fig:s4}, and compare the obtained spectrum with a system with only NN couplings in Fig.~\ref{fig:s6}. For fluxes close to $\pi$, where the geometrical angle is large and thus the NNN distances relatively small, the deformation of the bands is evident. In Fig.~\ref{fig:s6}(a), one can observe that the top and bottom bands are not flat, but rather dispersive, as is a sizeable portion of the central band. Conversely, for $\phi = \pi/2$ and lower, the deviations caused by these couplings become much smaller and increasingly negligible. Therefore, we limit the study to $0<\theta\leq\frac{\pi}{4}$ (corresponding to $0<\phi\leq\frac{\pi}{2}$), where only the NN couplings mentioned in the main text play a significant role.


\bibliography{biblio}

\end{document}